\documentclass[aps,prb,twocolumn,superscriptaddress]{revtex4}

\usepackage{epsfig}

\begin{document}

\title{Optimum pinning of the vortex lattice in extremely type-II layered
superconductors}
\author{C.~E.~Creffield}
\affiliation{Instituto de Ciencia de Materiales (CSIC), Cantoblanco, E-28049
Madrid, Spain}
\thanks{Current address: Dipartimento di Fisica, Universit\`a di Roma
``La Sapienza'', Piazzale Aldo Moro 2, I-00185 Roma, Italy}
\author{J.~P.~Rodriguez}
\affiliation{Dept. of Physics and Astronomy, California State University,
Los Angeles, California 90032, USA}

\date{\today}

\begin{abstract}
The two-dimensional (2D) vortex lattice in the extreme type-II limit is studied
by Monte Carlo simulation of the corresponding 2D Coulomb gas, with
identical pins placed at sites coinciding with the zero-temperature triangular
vortex lattice. At weak pinning we find evidence for 2D melting into an
intermediate hexatic phase. The strong pinning regime shows a
Kosterlitz-Thouless transition, driven by interstitial
vortex/anti-vortex excitations. A stack of such identical layers with a weak
Josephson coupling models a layered superconductor
with a triangular arrangement of columnar pins at the matching field.
A partial duality analysis finds that
layer decoupling of the flux-line lattice does not occur
at weak pinning for temperatures below 2D melting.
\end{abstract}

\maketitle

\section{Introduction}

It well known that the motion of vortex lines in the mixed
phase of a type-II superconductor generates dissipation,
and hence that an unpinned vortex-lattice state is
in fact resistive \cite{tinkham}. This has been confirmed recently
in the mixed phase of clean high-temperature superconductors,
where the superconductivity of samples with a strip
geometry is found to be very much superior to that in samples with
a Corbino disk geometry \cite{fuchs}. Surface barriers in the
strip geometry prohibit rigid motion of the vortex lattice,
while the Corbino disk geometry allows for rigid rotations of
the vortex lattice.
The above phenomena can be understood theoretically in the extreme
type-II limit, where magnetic screening is absent \cite{tinkham}.
Rigid translations of the vortex lattice result in an
infrared divergence that destroys phase coherence at any
temperature \cite{maki-moore}.
This infrared divergence can be removed by excluding rigid motion
of the vortex lattice through surface barriers, whereupon phase coherence
is restored \cite{jpr01}.

Defects in the bulk of a superconductor can also effectively
prohibit the rigid motion of the vortex lattice \cite{tinkham}.
In this work we study the nature of phase coherence in
an extremely type-II layered superconductor, with
magnetic field oriented perpendicular to the layers,
and containing an array of correlated pins.
We choose to use an optimum arrangement of
identical columnar pins \cite{blatter},
the locations of which match the triangular
vortex lattice at zero temperature.
Such a configuration can be realized experimentally  by
artificial ``anti-dot'' arrays \cite{antidots}.
The Josephson coupling between layers is turned off initially,
thus allowing us to model the system of vortices in each layer
by a two-dimensional (2D) Coulomb gas with a uniform charge
background and a commensurate pinning potential. (cf. ref. 7)
We employ Monte Carlo (MC) simulations to uncover the thermodynamic
phase diagram of this system under periodic boundary conditions.
The depth of the pinning potential,
$U$, becomes a useful control parameter.
As $U$ increases from zero, the ``floating'' vortex lattice
phase \cite{hattel_xy,franz_prb}
that exists at $U = 0$ becomes pinned at a critical $U_p$.
A finite-size analysis shows that $U_p$ tends to zero in the
thermodynamic limit, and that floating is prohibited initially by
a sparse distribution of pinned vortices.
Phase coherence is then restored at a yet stronger pinning, $U_m$.
The vortex lattice inside the range
$U_p < U < U_m$ therefore shows no phase coherence
despite the fact that it does not float!
The identification of this regime
with the {\em hexatic} phase \cite{HN} is indicated
by recent theoretical work \cite{jpr01}.
Indeed, a modest size analysis demonstrates that
this phase shows (strict)
long-range orientational order,
%typical configurations of the vortex lattice contain
%unbound dislocations in this regime, 
which is a hallmark of the hexatic phase \cite{HN}.
The strong-pinning regime at the other extreme
exhibits a standard Kosterlitz-Thouless (KT) transition
driven by the unbinding of vortex/anti-vortex pairs
that are {\em not} linked to the vortex lattice \cite{KT},
which is now fixed to the pins and appears to be irrelevant \cite{jpr00}.
Lastly, the effects of a weak Josephson coupling between layers
is determined through the application of a partial duality analysis of the
corresponding layered $XY$ model \cite{jpr00}.
On this basis, we conclude that the superconducting-normal transition
shown by such an optimally pinned vortex lattice lies within the universality
class of the three-dimensional (3D)
$XY$ model.

\section{2D vortex lattice with commensurate pins}

Consider an infinite stack of isolated superconducting layers
in a perpendicular external magnetic field.
Each layer is assumed to be identical in order
to reflect the correlated pinning.
A weak Josephson coupling will be switched on later.
Magnetic screening effects can be neglected in the extreme
type-II limit assumed throughout,
in which case the $XY$ model over the square lattice with
uniform frustration provides a qualitatively
correct description of the mixed phase of each layer.
The corresponding Boltzmann distribution is set by the
energy functional
\begin{equation}
  E_{XY}^{(2)} = - \sum_{\mu = x, y} \sum_{\vec r} J_{\mu} (\vec r)\,
{\rm cos} [\Delta_{\mu} \phi (\vec r) - A_{\mu} (\vec r)]
\label{2DXY}
\end{equation}
for the superfluid kinetic energy in terms of the superconducting phase
$\phi (\vec r)$.
Here $\Delta_{\mu} \phi (\vec r) =
\phi(\vec r + a \hat\mu) - \phi(\vec r)$ and
$\vec A = (0, 2\pi f x/ a)$ make up the local supercurrent,
where $f$ denotes the concentration of vortices over the
square lattice, with lattice constant $a$.
The local phase rigidity $J_{\mu} (\vec r)$
is assumed to be constant over most of the
nearest-neighbor links $(\vec r, \vec r + a \hat\mu )$,
with the exception of those links in the vicinity of a pinning site.
We shall next take the Villain approximation
which is generally valid at low temperature \cite{jose,ID}.
After making a series of standard manipulations, we obtain
a Coulomb gas ensemble with pins    that
describes   the vortex degrees of freedom on the dual square lattice.
The ensemble is weighted by the Boltzmann
distribution set by the energy functional
\begin{widetext}
\begin{equation}
  E_{\rm vx} =	(2\pi)^2   \sum_{(\vec R_1, \vec R_2)}
[Q(\vec R_1) - f]\ J G^{(2)} (\vec R_1, \vec R_2)\ [Q(\vec R_2) - f]
+  \sum_{\vec R} V_p (\vec R) \, | Q (\vec R) |^2
 \ ,
\label{ham}
\end{equation}
\end{widetext}
in terms of the integer vorticity field $Q (\vec R)$ over
the sites $\vec R$ of the dual lattice.
The logarithmic interaction between the vortices can be
expressed as a Greens function
$J G^{(2)} = \sum_n | n \rangle \varepsilon_n \langle n |$,
where the states $| n \rangle$ diagonalize the operator
$-\sum [J_x^{-1}  \Delta_y^2 + J_y^{-1}	 \Delta_x^2]$
with corresponding eigenvalues $\varepsilon_n^{-1}$.
Here, the lattice difference operator $\Delta_{y (x)}$ acts
between those adjacent points on the dual lattice that
are split by the link
%$(\vec r, \vec r + a \hat x (\hat y)  )$
on the base lattice that the coupling constant $J_{x (y)}$ refers to.
The effective coupling energy $J$ is set by the
requirement that
$\langle \vec q_1 | G^{(2)} | \vec q_2 \rangle	= q_1^{-2} a^{-2} \delta_{1,2}$
for plane waves $| \vec q_i \rangle$ in the long wavelength
limit, $q_i\rightarrow 0$.
The pinning potential, on the other hand, originates from the
contribution to the former Greens function
by bound states localized at the pinning sites:
$V_p (\vec R) = (2\pi)^2 \sum_{\varepsilon_n < 0}
\varepsilon_n | \langle \vec R | n \rangle |^2$.
Finally it is instructive to point out that the phase
rigidity can also be directly computed within the Villain
approximation, where it is given by \cite{minnhagen} one over
the dielectric constant of the Coulomb gas
ensemble (\ref{ham}).
In particular, for an $L \times L$ square
mesh with periodic boundary conditions, we have
\begin{equation}
\rho_s / J    =
 1 - \lim_{k \rightarrow 0}(2 \pi)^2 \beta
\left (\langle Q_{\vec k} Q_{-\vec k} \rangle
-\langle Q_{\vec k}\rangle \langle Q_{-\vec k} \rangle \right )
 /k^2 a^2 L^2  \ ,
\label{epsinv}
\end{equation}
where $\beta = J/k_B T$ is the inverse temperature of the system,
and $Q_{\vec k} = \sum_{\vec R} Q (\vec R) e^{i \vec k\cdot \vec R}$
is the Fourier transform of the charge density.

To proceed further, we shall first replace the Greens function mediating
the interaction between vortices in Eq. (\ref{epsinv}) with
the standard one $G^{(2)} = - \nabla^{-2}$ over the square lattice
subject to periodic boundary conditions \cite{jose,ID}.
This approximation neglects only the short-range features of the
interaction energy near the pinning centers.
Secondly, we shall consider the optimum pinning configuration:
$V_p (\vec R) = -U$ for points
$\vec R$ that coincide with the triangular
vortex lattice at zero temperature,
and $V_p (\vec R) = 0$ otherwise.
The long-range logarithmic interaction among vortices
enforces charge neutrality with the uniform background charge density
$f$, such that:
\begin{equation}
\sum_{\vec R}  Q (\vec R)  = f L^2 .
\label{charge_neut}
\end{equation}
This means that the system of vortices is incompressible
at all temperatures. Vacancies and interstitials are therefore
impossible at long wavelengths.
In the absence of extrinsic pins\cite{hattel_xy}$^,$\cite{franz_prb}, 
$U = 0$, and at low vorticity,
$f < 1/30$, the triangular vortex-lattice
depins from the underlying square lattice at a temperature
$k_B T_p^{(0)} = 1.5  f J$. At higher temperatures  it ``floats'', before
melting at $k_B T_{m}^{(0)} = J / 20$.
At low vorticity and low temperature, the difference
in the internal	 energy between the floating and the pinned
vortex lattice phases is therefore
$\Delta E  = N_{\rm vx-p} U$, where $N_{\rm vx-p}$ denotes the
average number of pinned vortices.
The corresponding difference in the entropy is
$\Delta S  =  k_B \, {\rm ln} (2/f)$, which is not extensive.
The balance in free energy, $\Delta E = T \Delta S$,
yields a (first-order) transition between the
two phases at a critical pinning strength
\begin{equation}
U_p = N_{\rm vx-p}^{-1} k_B T \,  {\rm ln} (2/f)
\label{Up}
\end{equation}
that vanishes in the thermodynamic limit, $L\rightarrow\infty$.
Below we shall confirm this prediction at weak pinning,
as well as exploring the effect of strong pinning through MC
simulation of the 2D Coulomb gas ensemble (\ref{ham}).

The following thermodynamic quantities are measured in the MC
simulations of the Coulomb gas ensemble (\ref{ham}) describing
a 2D vortex lattice with optimum pins.
Phase coherence is probed by the superfluid stiffness (\ref{epsinv}).
Pinning of the vortex lattice is determined by checking
for the appearance of Bragg peaks in the vortex density,
$S_0(\vec k) = |\langle Q_{\vec k}\rangle|^2$.
This is quantified by taking the ratio between the heights of the
first-order and zero-order Bragg peaks, which we term the ``Bragg ratio''.
{\em Intrinsic} positional correlations among the vortices,
on the other hand,
are measured through the fluctuation contribution to the structure function,
$S_1(\vec k) = \langle Q_{\vec k} Q_{-\vec k}\rangle
	       -  \langle Q_{\vec k}\rangle \langle Q_{-\vec k}\rangle $.
Lastly we also measure
the six-fold orientational order parameter characteristic of the
triangular vortex lattice:\cite{hattel_xy}$^-$\cite{HN}
$\phi_6 = N_{\rm vx}^{-2} \sum_{j, k}
\langle {\rm  exp}[ 6 i (\theta_j - \theta_k)] \rangle $.
Here $\theta_j$ is the angle between a fixed direction in space and the
direction of the bond between the $j$-th vortex and its nearest neighbor.
In general, this orientational order parameter
decays algebraically with system size as
$L^{-\eta_6}$ in the thermodynamic limit, $L\rightarrow\infty$.
The correlation exponent $\eta_6$ is null in the case
of a  2D vortex lattice with strict long-range order,
while  $\eta_6 = 2$ if only short-range orientational order
exists\cite {franz_prb}.\\
%algebraicly with size $L$  in both the hexatic and liquid phases

The MC scheme used closely follows that developed by Lee and
Teitel \cite{teitel_prb}.
A single MC move consists of selecting a lattice
point and one of its nearest neighbors at random,
and adding a unit charge
to one of the points and subtracting unit charge from the other, thereby
keeping the total charge of the system constant. This alteration is then
either accepted or rejected according to
the standard Metropolis algorithm.
At temperatures lower than $J/4 k_B$, simulations revealed
that the accepted configurations
{\em only} possessed charges of $Q = + 1$,
the total number of charges, $N_{\rm vx}$, being set
by overall charge neutrality (\ref{charge_neut}).
All other configurations, such as those with multiple or
negative charges, were heavily penalized on energetic grounds.
This permits a much less computationally costly MC move to be employed
in this regime, which consists of selecting one of these $N_{\rm vx}$
charges at random and moving it to an unoccupied neighboring lattice site.
Extensive simulations at low temperatures were run
using both MC updating methods to confirm that they indeed gave the same
results, and thereafter the second algorithm was used to obtain the bulk
of the low-temperature results in this paper.
The MC estimates are based on between 8000
and 12000 measurements, following an equilibration from a random
initial configuration consisting of 8000 MC sweeps. In all cases this
process was repeated a number of times, using a different initial state,
to ensure that the simulations were not becoming ergodically trapped.

\begin{figure}
\centerline{\epsfxsize=60mm \epsfbox{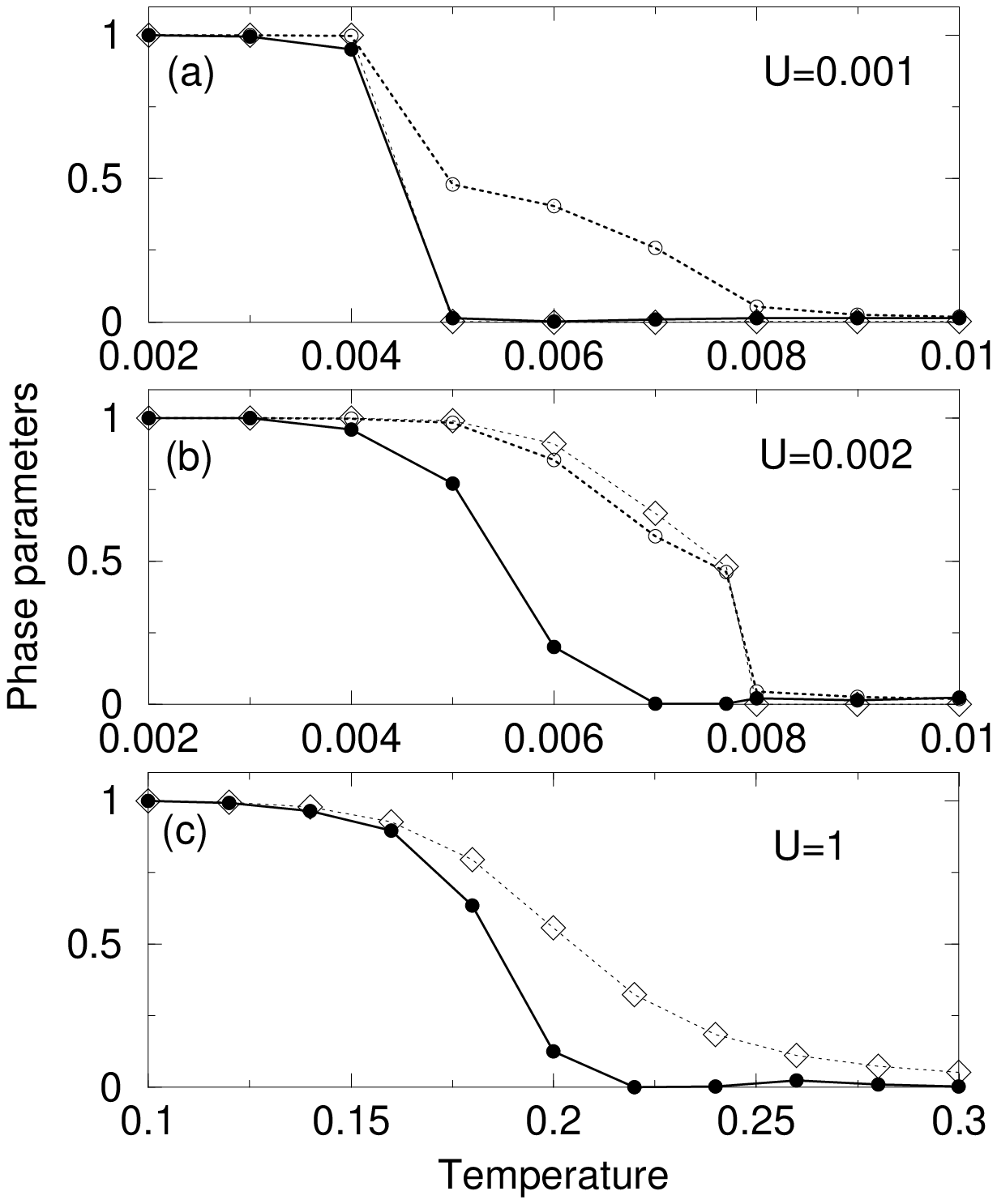}}
\caption{Phase transitions at $f=56^{-1}$ on a	$112 \times 112$ lattice,
for three values of pinning potential. Thick solid line
(black dots) = $\rho_s / J$,
thick dotted line (open circles) = $\phi_6$,
thin dotted line (open diamonds) = Bragg ratio
in $S_0 (\vec k)$. Error bars are smaller than symbols.
All energies are given in units of $2\pi J$.}
\label{transitions}
\end{figure}

Fig. \ref{transitions} displays temperature
profiles of the various physical probes
that were listed above, for three different regimes
of the strength of the pinning potential.
The phase-incoherent floating phase lies in between the pinned vortex
lattice and the vortex liquid as a function of temperature in the
no-pinning regime shown in Fig. \ref{transitions}a
\cite{franz_prb,hattel_xy}.
A strange phase-incoherent state that does {\em not} float,
on the other hand, lies in between the conventional solid and liquid phases
in the weak-pinning regime (Fig. \ref{transitions}b).
Table \ref{fss} demonstrates that this observation
is not a spurious  size effect.	 Notice, in particular,
how the orientational order parameter $\phi_6$
and the magnitude of the first-order Bragg peak saturate with increasing size,
and how the phase rigidity remains null throughout.
The former implies that the ``strange'' vortex phase exhibits
{\it strict} long-range orientational order, with a correlation exponent
$\eta_6 = 0$.\cite{franz_prb}
Fig. \ref{diagram} displays the resulting phase diagram in the $U$-$T$ plane.
The boundary, $U_p(T)$,	that separates the floating phase (I)
and the ``strange'' phase (II) extrapolates to zero roughly as $L^{-1}$
at fixed temperature. This was determined from MC simulations
at a temperature $k_B T / 2\pi J =0.0060$ for
two different sizes, $L = 56  \ \mbox{and} \ 112$, with
the addition of the thermodynamic limit, $L\rightarrow\infty$, 
under the assumption that $U_p = 0$ there.
(Note that our MC simulations show
metastability in the vicinity of this first-order pinning transition,
which reflects the two-fold orientational degeneracy of
the floating states.)
Consistent with the simple balance of free energy (\ref{Up}),
we then conclude that the floating phase exists only
in the absence of extrinsic pins ($U = 0$) in the thermodynamic limit.
The last size analysis also implies by Eq. (\ref{Up}), 
that the number of pinned vortices in the ``strange'' phase 
scales as $N_{\rm vx-p}\propto L$, which is {\it subthermodynamic}. 
Finally, at yet stronger pinning, the line $U = U_m (T)$
along which macroscopic phase coherence sets in shows only
minor size dependence for the three $L \times L$ lattices 
that we simulated.

\begin{table}
\begin{center}
\begin{tabular}{|l|c|c|c|}
\hline
 $L$ &	$\rho_s/J$ & Bragg ratio & $\phi_6$ \\
\hline
56  & $0.001 \ \pm \ 0.002$ & $0.329 \ \pm \ 0.061$ & $0.603 \ \pm \ 0.008$ \\
112 & $0.000 \ \pm \ 0.001$ & $0.765 \ \pm \ 0.003$ & $0.677 \ \pm \ 0.002$ \\
168 & $0.001 \ \pm \ 0.001$ & $0.784 \ \pm \ 0.003$ & $0.690 \ \pm \ 0.003$ \\
\hline
\end{tabular}
\caption{Finite size study of the Coulomb gas at a
temperature $k_B T / 2\pi J =0.0065$
 and pinning potential $U / 2\pi J = 0.002$
inside the putative hexatic phase.
This strength of pinning is slightly larger than $U_p$ for $L = 56$,
which explains the relatively small Bragg ratio at this size
(recall that $U_p$ decreases with size).}
\label{fss}
\end{center}
\end{table}

A recent theoretical analysis of the 2D Coulomb gas (\ref{ham})
in the absence of bulk pinning finds that an intermediate
{\em hexatic} phase can indeed exist if rigid translations of the
vortex-lattice are prohibited \cite{jpr01}.
This phase contains unbound dislocations that generate
appreciable fluctuations in the center of mass of the
2D vortex lattice.
These fluctuations are responsible for
both the destruction of macroscopic
shear rigidity and of macroscopic phase coherence
at a   2D melting transition.
The hexatic phase
contrasts with the conventional pinned vortex-lattice phase
that shows phase coherence.
It also differs from the vortex liquid
phase by the presence of orientational order \cite{HN}.
We propose to identify the strange intermediate phase (II)
that neither floats nor	   shows macroscopic phase coherence,
but that exhibits strict  long-range orientational order
(see Figs. \ref{transitions} and \ref{diagram}, and Table \ref{fss}),
with such a hexatic phase.
Fig. \ref{structure} displays the fluctuation part of the structure
function, $S_1 (\vec k)$, in the putative hexatic phase
found in the weak-pinning regime.
It shows six-fold
symmetric Bragg peaks of low order, which strikingly resemble those
obtained experimentally in real-life hexatic phases\cite{hexexp}.
Further, Fig. \ref{dislocation} shows
a typical configuration of the vortex lattice
in the putative hexatic phase near the pinning threshold.
We may note the presence of an unbound dislocation and the small fraction
of vortices that are pinned.
The strict long-range orientational order (Fig. \ref{transitions}b),
the  six-fold pattern shown by this intrinsic structure function
(Fig. \ref{structure}),
and the presence of unbound dislocations (Fig. \ref{dislocation})
are all consistent with the
identification of the ``strange'' vortex lattice
with a hexatic phase over a commensurate substrate \cite{HN}.

The vortex lattice tends to be fixed to each and
every commensurate pin at strong
$U/2\pi J > 0.01$ (see Fig. \ref{diagram}).
Interstitial vortex/anti-vortex excitations
are the only remaining degrees of freedom
in such case.
The temperature dependence of the phase rigidity (\ref{epsinv}) obtained
from our MC simulations strongly resembles that of
the zero-field case ($f = 0$) with no
extrinsic pins ($U = 0$) \cite{franz_prb}.
In particular, Fig. \ref{transitions}c
shows how  $\rho_s$ decreases smoothly from $J$ to $0$
at the expected KT transition temperature \cite{KT},
$k_B T_c^{(0)} \cong {\pi\over 2} J$.
The vortex lattice subsequently	 frees itself from the extrinsic pins at a
higher temperature, $T_p$, as shown by
Figs. \ref {transitions} and \ref{diagram}.

\section{3D vortex lattice with optimum columnar pins}

We finally apply the above results to the question of
phase coherence in
the vortex-lattice phase of layered superconductors.
In the extreme type-II limit, this
system can be modeled by an infinite stack of
$XY$-model layers with uniform frustration, Eq. (\ref{2DXY}),
but with an additional  Josephson coupling,
$J_z = J/\gamma^{\prime 2}$,
between all nearest-neighbors across adjacent layers.
Also, in this limit
the magnetic coupling to vortices in adjacent layers
can be accounted for by weak optimum columnar pinning, $V_p (\vec R)$,
within the ``substrate potential'' approximation\cite{substrate}.
Such a layered $XY$ model can be analyzed through
a partial duality transformation that is ideally suited to the
weak-coupling limit.
This leads to the following partition function that encodes
the thermodynamics of the coupled system:
$Z_{\rm CG} = \sum_{\{ n_z  \}}
(\beta / 2\gamma^{\prime 2})^{N[n_z]}
\Pi_l C [p_l]$,
where $n_z (\vec r, l)$
is an integer link field on 2D points $\vec r$ between
adjacent layers $l$ and $l+1$ \cite{jpr00}. Here
$C [p_l] =
\langle{\rm exp} [ i \sum p_l(\vec r) \phi (\vec r, l)]\rangle_{J_z =0}$
is the phase auto-correlation function of an isolated layer $l$
probed at the dual charge that collects onto that layer:
$p_l (\vec r) = n_z (\vec r, l-1) - n_z (\vec r, l)$.
Also, $N [n_z]$ counts the total number of dual charges, $n_z = \pm 1$.
The latter system is dilute in the weak-coupling limit
reached at large model anisotropy parameters,
$\gamma^{\prime}\rightarrow\infty$.
It has been shown recently by one of us \cite{jpr01}
that the phase auto-correlator for
a pure 2D vortex lattice that cannot move rigidly
has the form
$|C (1,2)| = (\rho_s/J) (r_{0} / r_{12})^{\eta_{2D}}$,
with a small 2D correlation exponent, $\eta_{2D} < (28\pi)^{-1}$.
Here $r_0$ is of the order of the inter-vortex spacing
and $r_{12}$ denotes the separation between the two probes.
Since commensurate pinning, $V_p (\vec R)$, 
increases phase coherence (see Fig. \ref{transitions}),
the previous bound on $\eta_{2D}$ continues to hold.
Yet the application of the above duality analysis yields a phase rigidity
across layers equal to\cite{jpr00}
$\rho_{s}^{\perp}  = (\rho_s/\gamma^{\prime 2})
 (r_0 /\gamma^{\prime} a)^{\eta_{2D}}$.
The extremely small bound on $\eta_{2D}$ then implies that a decoupling
crossover, $\rho_{s}^{\perp} \ll  \rho_s/\gamma^{\prime 2}$,
occurs only for astronomically large anisotropy,
$f\gamma^{\prime 2} > 10^{38}$.
This indicates that layer decoupling does not occur in practice in
the vortex-lattice phase of extremely type-II layered superconductors
at temperatures below 2D melting (cf. ref.  7).

The above partial duality analysis of
course also applies directly to the
question of optimum columnar
pinning in the vortex-lattice phase of
strongly type-II layered superconductors.
In the strong pinning regime,
Fig.  \ref {transitions} c,
our MC simulation results for a single layer
find conclusive evidence for
a standard KT phase transition driven by
the unbinding of interstitial vortex/anti-vortex pairs.
This implies that the autocorrelation functions $C [p]$
that appear in $Z_{\rm CG}$
are precisely those corresponding to the 2D $XY$ model in the
absence of frustration ($f = 0$) and extrinsic	pinning ($U = 0$),
up to a gauge transformation.
We thereby conclude that the strongly pinned vortex lattice at the
matching field goes through a superconducting/normal transition
that is second-order, and that coincides with the
universality class of the standard 3D $XY$ model.

\section{Discussion and conclusions}

\begin{figure}
\centerline{\epsfxsize=60mm \epsfbox{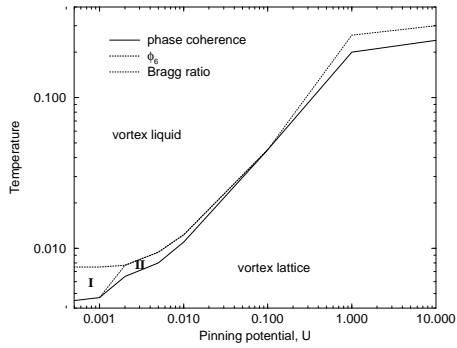}}
\caption{Phase diagram for $f=56^{-1}$ on a  $112 \times 112$ lattice.
Roman numerals I and II indicate the floating and the hexatic phases,
respectively.
All energies are given in units of $2\pi J$.}
\label{diagram}
\end{figure}

\begin{figure}
\centerline{\epsfxsize=60mm \epsfbox{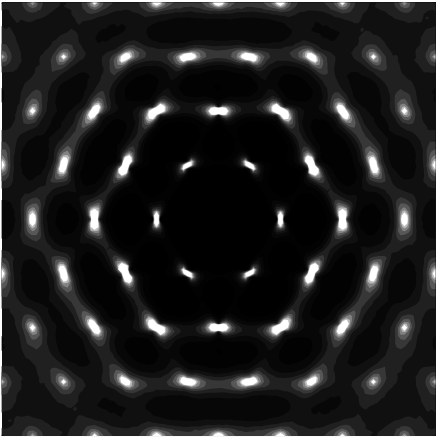}}
\caption{Intrinsic structure function, $S_1 (\vec k)$,
 for the hexatic phase (II) at
$k_B T/ 2\pi J = 0.007$ and  $U/2\pi J = 0.002$.}
\label{structure}
\end{figure}

\begin{figure}
\centerline{\epsfxsize=50mm \epsfysize=50mm \epsfbox{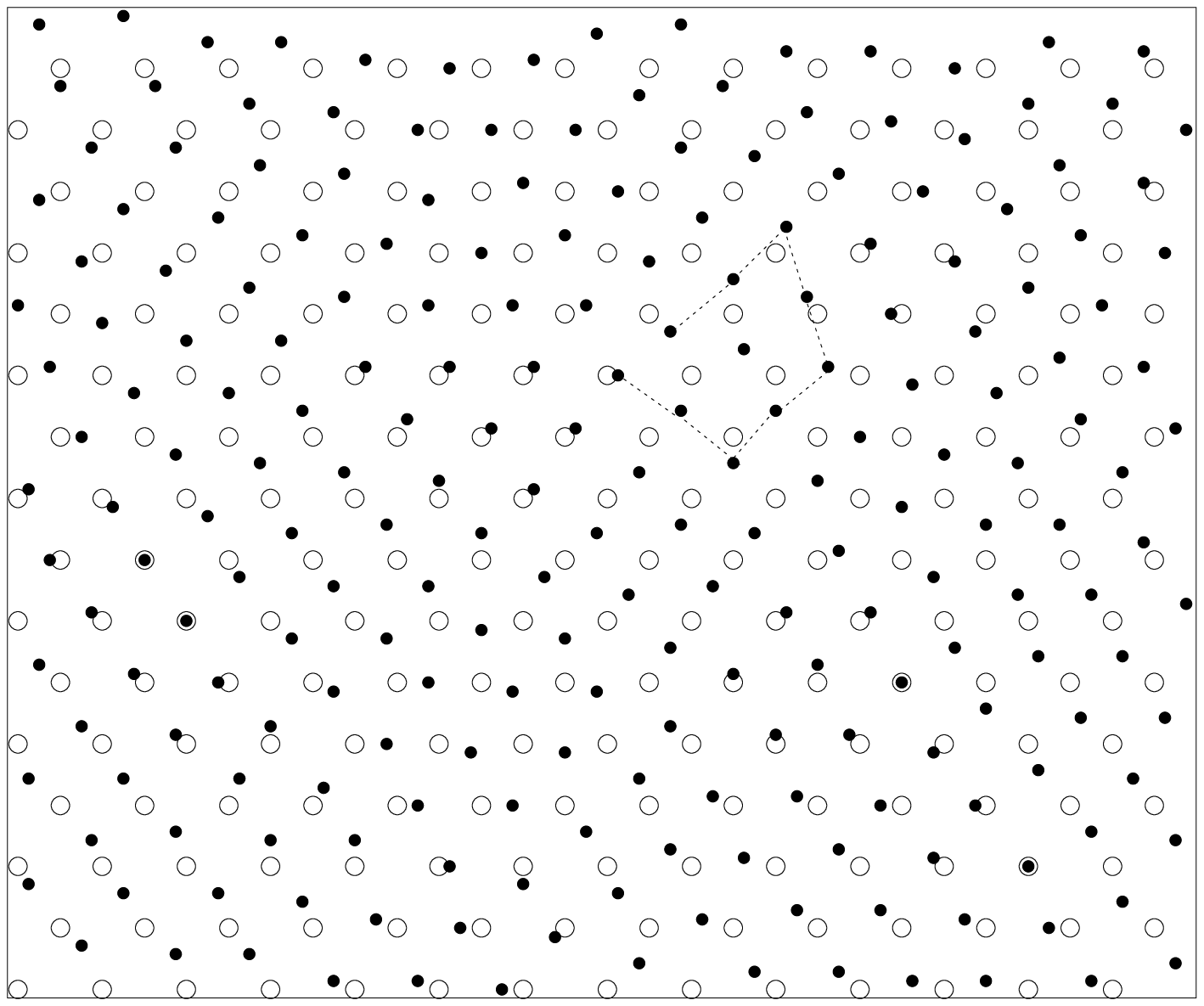}}
\caption{Typical configuration (black dots)
for the hexatic phase (II) at $f=56^{-1}$ on a	$112 \times 112$ lattice,
at a temperature of $k_B T / 2\pi J = 0.0067$ and pinning potential
(open circles) $U / 2\pi J = 0.0008$.
Observe the presence of an unbound dislocation inside the Burgers
circuit shown, and the absence of disclinations.
Also notice the small number of pinned vortices
(dot in circle).}
\label{dislocation}
\end{figure}

In summary, we have studied the nature of phase coherence
in the 2D vortex lattice at the extreme type-II limit
with identical commensurate pins through Monte Carlo
simulation of the corresponding 2D Coulomb gas, Eq. (\ref{ham}).
Our most striking result is the identification of a
strange intermediate phase (II) that is pinned, but that
shows no macroscopic phase coherence.
It lies in the midst of a floating vortex lattice,
a pinned vortex lattice, and a vortex liquid phase
(see  Fig. \ref{diagram}). A recent calculation
has indicated that an intermediate hexatic phase
with such characteristics is indeed expected in the
2D vortex lattice at the extreme type-II limit (\ref{2DXY}) if rigid
translations of the vortex lattice are prohibited \cite{jpr01}.
The present calculations show that an arbitrarily weak
array of commensurate pins have this effect in the thermodynamic
limit [see Eq. (\ref{Up}) and Fig. \ref{diagram}].
We have proposed, in conclusion, that the
strange intermediate phase just mention is in fact
a hexatic vortex lattice.

On the contrary, theoretical calculations of 2D lattice melting in the
presence of a weak commensurate substrate potential
do {\em not} predict the existence of
separate liquid and hexatic phases \cite{HN}.
It is quite possible, then, that the transition
that we observe through Monte Carlo simulation
between a  hexatic vortex lattice (II) that is pinned
and a vortex liquid that is not pinned
is in fact a sharp cross-over transition.
In such case, the phase diagram shown in Fig. \ref{diagram}
would be topologically equivalent to that predicted
by theory (see Ref. \cite{HN}, Fig. 4, right-hand side).
More extensive Monte Carlo simulations may be needed
to settle this question.

\acknowledgments

The authors thank R. Markiewicz, F. Nori, F. Guinea,
R. Mulet and E. Altshuler  for valuable discussions.
CEC acknowledges support from the EU TMR programme.
JPR acknowledges the hospitality of the
Instituto de Ciencia de Materiales de Madrid,
where this work was initiated.

\begin{thebibliography}{}
\bibitem{tinkham}
{M. Tinkham, {\it Introduction to Superconductivity}
(McGraw-Hill, New York, 1996) Chap. 5.}

\bibitem{fuchs}
{S.F.W.R. Rycroft, R.A. Doyle, D.T. Fuchs, E. Zeldov, R.J. Drost,
P.H. Kes, T. Tamegai, S. Ooi and D.T. Foord,
Phys. Rev. B {\bf 60}, R757 (1999).}

\bibitem{maki-moore}
{K. Maki and H. Takayama, Prog. Theor. Phys. {\bf 46}, 1651 (1971);
M.A. Moore, Phys. Rev. B {\bf 45}, 7336 (1992).}

\bibitem{jpr01}
{J.P. Rodriguez, Phys. Rev. Lett. {\bf 87}, 207001 (2001).}

\bibitem{blatter}
{G. Blatter, M.V. Feigel'man, V.B. Geshkenbein, A.I. Larkin, and V.M. Vinokur,
Rev. Mod. Phys. {\bf 66}, 1125 (1994).}

\bibitem{antidots} K. Harada, O. Kamimura, H. Kasai, T. Matsuda,
A. Tonomura and V.V. Moshchalkov, Science {\bf 274}, 1167 (1996).

\bibitem {substrate}
{M.J.W. Dodgson, V.B. Geshkenbein and G. Blatter,
Phys. Rev. Lett. {\bf 83}, 5358 (1999).}

\bibitem{hattel_xy}
{S{\o}ren A. Hattel and J.M. Wheatley, Phys. Rev. B {\bf 51}, 11951 (1995).}

\bibitem{franz_prb}
{M. Franz and S. Teitel, Phys. Rev. B {\bf 51}, 6551 (1995).}

\bibitem{HN}
{D.R. Nelson and B.I. Halperin, Phys. Rev. B {\bf 19}, 2457 (1979).}

\bibitem{KT}
{J.M. Kosterlitz and D.J. Thouless, J. Phys. C {\bf 6}, 1181 (1973).}

\bibitem{jpr00}
{J.P. Rodriguez, Phys. Rev. B {\bf 62}, 9117 (2000);
Physica C {\bf 332}, 343 (2000); Europhys. Lett. {\bf 54}, 793 (2001).}

\bibitem{jose}
{J.V. Jos\' e, L.P. Kadanoff, S. Kirkpatrick and
D.R. Nelson, Phys. Rev. B {\bf 16}, 1217 (1977).}

\bibitem{ID}
{C. Itzykson and J.  Drouffe, {\it Statistical Field Theory},
Vol. 1, (Cambridge Univ.  Press, Cambridge, 1991) chap. 4.}

\bibitem{minnhagen}
{P. Minnhagen and G.G. Warren, Phys. Rev. B {\bf 24}, 2526 (1981);
P. Minnhagen, Rev. Mod. Phys. {\bf 59}, 1001 (1987).}

\bibitem{teitel_prb}
{Jong-Rim Lee and S. Teitel, Phys. Rev. B {\bf 46}, 3247 (1992).}

\bibitem{hexexp}
{R. Seshadri and R.M. Westervelt, Phys. Rev. Lett. {\bf 66}, 2774 (1991);
C-F. Chou, A.J. Jin, S.W. Hui, C.C. Huang and J.T. Ho,
Science, {\bf 280}, 1424 (1998).}

\end{thebibliography}
\end{document}